\documentclass[prd, aps, superscriptaddress, preprintnumbers, twocolumn, floatfix, nofootinbib]{revtex4}
\pdfoutput=1

\usepackage{amsfonts}
\usepackage{amsmath}
\usepackage{amssymb}
\usepackage{bm}
\usepackage{dcolumn}
\usepackage{graphicx}   
\usepackage[latin1]{inputenc}
\usepackage{latexsym}
\usepackage{rotating}
\usepackage{hyperref}
\usepackage{graphicx}
\usepackage{color}

\newcommand\be{\begin{equation}}
\newcommand\ba{\begin{eqnarray}}
\newcommand\ee{\end{equation}}
\newcommand\ea{\end{eqnarray}}

\begin{document}

\title{Creating Spatial Flatness by Combining String Gas Cosmology and Power Law Inflation}

\author{Vahid Kamali}
\email{vkamali@ipm.ir}
\affiliation{Department of Physics, McGill University, Montr\'{e}al, QC, H3A 2T8, Canada\\
Department of Physics, Bu-Ali Sina (Avicenna) University, Hamedan 65178, 016016, Iran\\
School of Physics, Insitute for Research in Fundamental Sciences (IPM), 19538-33511, Tahran, Iran}

\author{Robert Brandenberger}
\email{rhb@physics.mcgill.ca}
\affiliation{Department of Physics, McGill University, Montr\'{e}al, QC, H3A 2T8, Canada}

\date{\today}

%%%%%%%%%%%%%%%%%%%%%%%%%%%%%%%%%%%%%%%%%%%%%%%%%%%%%%%%%%%%%%%%%%%%%%%%%%%%%%%%%%%%%%%%%%%%%%
\begin{abstract}

We show that it is possible to combine an early phase of String Gas Cosmology which can explain the origin of the observed structures on cosmological scales with a short later period of power law inflation which creates spatial flatness. The resulting model is consistent with the ``swampland criteria'' and the constraints coming from the {\it Trans-Planckian Censorship Conjecture}. Such a construction is not possible using only canonical slow-roll inflation, but it can emerge in the warm inflation scenario or in cold inflation with an exponential potential. The resulting cosmology is non-singular. We discuss the spectrum of cosmological perturbations resulting in our scenario. On large scales (scales which remain larger than the Hubble radius after the initial string gas phase) the spectrum is determined by the thermal string gas fluctuations set up in the primordial phase, and it is almost scale-invariant with a slight red tilt. On small scales, the perturbations produced during the inflationary phase dominate. On these scales, the spectrum is once again nearly scale-invariant. There is an intermediate range (scales which enter the Hubble radius during the period of inflation) where the string gas fluctuations are damped but continue to dominate over those produced during the period of inflation. On these scales the spectrum has a sharp red spectral index of $n_s - 1 \sim -2$.

\end{abstract}
%%%%%%%%%%%%%%%%%%%%%%%%%%%%%%%%%%%%%%%%%%%%%%%%%%%%%%%%%%%%%%%%%%%%%%%%%%%%%%%%%%%%%%%%%%%%%%

\maketitle

%%%%%%%%%%%%%%%%%%%%%%%%%%%%%%%%%%%%%%%%%%%%%%%%%%%%%%%%%%%%%%%%%%%%%%%%%%%%%%%%%%%%%%%%%%%%%%
\section{Introduction} 
\label{sec:intro}

{\it String Gas Cosmology} \cite{BV} (see also \cite{Perlt} for related work, and \cite{SGCrevs} for reviews) is an alternative to inflationary cosmology for producing an almost scale-invariant spectrum of cosmological perturbations \cite{NBV}, and also predicts a roughly scale-invariant spectrum of gravitational waves with a characteristic slight blue tilt \cite{BNPV}. String gas cosmology is based on considering the new degrees of freedom and new symmetries which differentiate string theory from point particle-based theories. In particular, String Gas Cosmology yields a nonsingular early universe scenario according to which there is an early phase in which the universe loiters in a state given by a hot gas of strings with temperature just below the Hagedorn value $T_H$ \cite{Hagedorn}, the maximal temperature which a gas of closed strings can take on when in thermal equilibrium.

While the horizon problem of Standard Big Bang cosmology is absent in String Gas Cosmology, the scenario does not provide a mechanism for producing the observed near spatial flatness of the current universe \footnote{Note that if String Gas Cosmology is implemented in the context of a bouncing cosmology, as proposed in \cite{Tirtho} and also suggested by considerations from Double Field Theory (see \cite{us}), then the flatness problem is trivially solved.}. Solving this {\it flatness problem} is one of the key successes of inflationary cosmology \cite{Guth}. Inflation also provides a mechanism of cosmological structure formation \cite{ChibMukh} based on quantum vacuum fluctuations which are stretched beyond the Hubble horizon by the accelerated expansion of space. 

The inflationary scenario of early universe cosmology is, however, tightly constrained by the {\it Trans-Planckian Censorship Conjecture} (TCC) \cite{BeVa} which states that no fluctuation modes whose physical wavelength was shorter than the Planck length should have ever exited the Hubble horizon and entered the classical domain. Demanding that inflation be consistent with the TCC and that scales which are currently probed in cosmic microwave background (CMB) experiments were sub-Hubble at the beginning of the period of inflation leads to the tight upper bound \cite{BBLV}
\be \label{Vbound}
V_{R}^{1/4} \, < \, 10^{10} {\rm GeV} \, 
\ee
on the energy scale at the end of inflation \footnote{To derive this bound, we assume almost constant Hubble expansion rate $H$ during inflation and immediate transition to the radiation phase of Standard Big Bang cosmology after inflation. Modified bounds obtained by relaxing the former condition are discussed in \cite{Kamali1}, modifications by assuming non-standard post-inflation cosmology were studied in \cite{mod1}, and changes when dropping the assumption of standard vacuum initial conditions for the fluctuations were investigated in \cite{Suddho}. The bound can be strengthened in the case that the inflationary phase is preceded by a non-accelerating phase described as in standard cosmology \cite{Edward}.}. Since it is the energy scale during inflation which sets the amplitude of the spectrum of primordial gravitational waves \cite{Starob}, the above upper bound on $V_R$ implies an upper bound on the tensor to scalar ratio $r$ of
\be \label{Rbound}
r \, < \, 10^{-30} \, ,
\ee
which in unobservably small given present technology of CMB cosmology. Note, in addition, that unnatural fine-tunings of the inflationary model are required in order to render (\ref{Vbound}) consistent with the observed amplitude of the spectrum of cosmological perturbations.

There is also a tension between standard {\it cold} inflation (obtained by using a canonically normalized and weakly coupled scalar matter field to obtain the almost exponential expansion of space in the context of Einstein gravity as the theory of space and time) and superstring theory. According to the {\it swampland conjectures} (see \cite{swamprevs} for reviews) the field range over which an effective theory of a scalar field $\varphi$ is consistent with underlying fundamental theory is constrained to be \cite{swamp1} smaller than $d m_{pl}$, where $m_{pl}$ is the Planck mass and $d$ is a constant of order 1, and its potential energy function $V(\varphi)$ for fields which dominate the evolution of the universe must obey \cite{swamp2} the condition
\be \label{Vcond}
\frac{|V^{\prime}|}{V} \, > \, \frac{c}{m_{pl}} \, ,
\ee
where $c$ is a constant of order 1, and a prime indicates a derivative with respect to $\varphi$ \footnote{For a determination of the value of the constant $c$ in the context of string gas cosmology see \cite{Samuel}.}. This condition is in conflict with the requirements of slow-roll inflation \cite{VS} (but not yet, given the current data, with the conditions on a quintessence potential for Dark Energy \cite{Lavinia}).

As shown in \cite{Kamali1}, in a model of power law inflation where $H$ decreases as $1/t$, the constraints from the TCC can be relaxed, and the swampland criterion on the slope of the potential can be satisfied \cite{Lavinia}. However, the spectrum of fluctuations emerging from quantum vacuum perturbations has a tilt which is too large to be compatible with observations.

Instead of trying to obtain cold power law inflation (inflation generated by a scalar field only very weakly coupled to other matter), it is possible to realize this scenario in the context of warm inflation. Warm inflation \cite{warm} in the high dissipation regime is a variant of scalar field-driven inflation in which the dynamics of the scalar field is dominated by friction due to coupling to other matter rather than Hubble friction. As a consequence, the potential $V(\varphi)$ can be sufficiently steep to meet the condition (\ref{Vcond}) without strongly violating the slow-roll conditions \cite{warm2}. The TCC, on the other hand, still imposes stringent constraints on the energy scale of warm inflation if we demand that inflation remains as a causal mechanism of explaining the observed structure on the largest scales \cite{warm3}.

In this Letter we point out that it is possible to combine an early phase of String Gas Cosmology which can explain the origin of the observed structures with a short later period of power law inflation which creates spatial flatness. The resulting model is consistent with the swampland criteria and the constraints coming from the TCC. We also show that this result cannot be achieved with a period of canonical slow-roll inflation. An advantage of the scenario we are proposing is that the amplitude of the spectrum of cosmological fluctuations and gravitational waves can be large without having to impose any unnatural fine-tunings of the inflationary model.

In the following we will use units in which the speed of light, Planck's constant and Boltzmann's constant are set to 1. The reduced Planck mass and Planck length are denoted by $m_{pl}$ and $l_{pl}$, respectively. We work in terms of the usual Friedmann-Lemaitre-Robertson-Walker metric with cosmological scale factor $a(t)$. The Hubble expansion rate is $H(t) \equiv {\dot{a}}/a$, the overdot denoting a time derivative. The inverse of $H(t)$ is the Hubble radius (Hubble horizon) $l_H(t)$ which separates scales (sub-Hubble) on which the cosmological fluctuations oscillate from scales (super-Hubble) where the oscillations freeze out, the fluctuations squeeze and can become classical (see e.g. \cite{MFB} for an in-depth review of the theory of cosmological fluctuations, \cite{RHBfluctsrev} for an overview, and \cite{Kiefer} for a discussion of the classicalization of fluctuations).

\section{Combining String Gas Cosmology with Power Law vs. Almost Exponential Inflation}

Figure 1 represents a sketch of the space-time diagram which we have in mind. The universe is assumed to begin as a hot gas of closed strings with temperature close to the limiting temperature $T_H$. We assume that a thermal equilibrium state is established. At a critical time $t_p$, a phase transition to an expanding phase of Standard Cosmology takes place. Since the phase transition at the end of the stringy phase results in the production of radiation, we assume that the universe is dominated by radiation after $t_p$. Between $t_i$ and $t_R$ we assume that there is a period of warm inflation. If the thermal correlation functions in the early stage have holographic scaling with size (i.e. they scale with the area as opposed to the volume), as they have for a gas of closed strings on a compact space \cite{NBV, Nayeri}, then the spectrum of cosmological fluctuations is scale-invariant with a slight red tilt, like what is predicted in simple inflationary models. The induced spectrum of gravitational waves is scale-invariant, but with a slight blue tilt \cite{BNPV}.

%\begin{widetext}
\begin{figure*}[t]
\begin{center}
\includegraphics[scale=1.2]{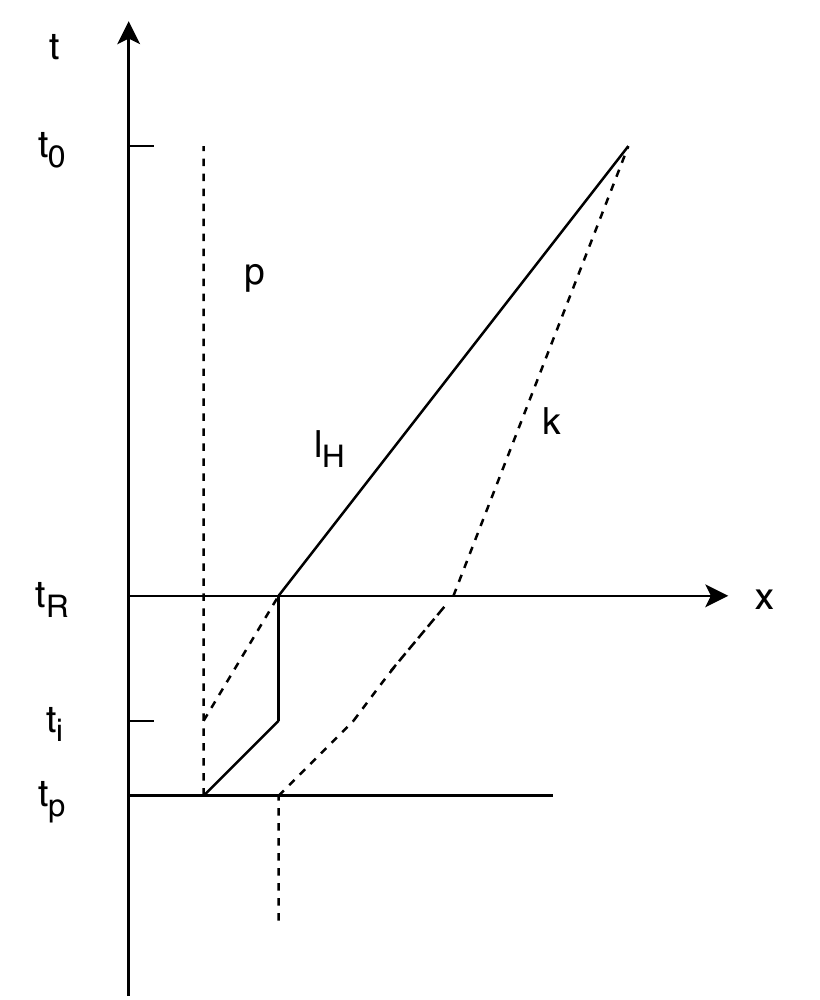}
\caption{Space-time (time along the vertical axis, physical distance along the horizontal axis) sketch of the cosmology of an early phase of String Gas Cosmology for $t < t_p$ with a later phase of inflation lasting from $t = t_i$ to $t = t_R$. We require that scales which are currently probed in CMB experiments (denoted by $k$ in the graph) are outside the Hubble horizon (denoted by $l_H$) at the beginning of inflation. In this case, the spectrum of cosmological perturbations and gravitational waves on these scales is set by the String Gas Cosmology phase. We demand that the inflationary phase does not violate the TCC. This condition is depicted to be marginally satisfied in the example shown in the sketch: the dashed line starting at $t_i$ with Planck length (the dashed vertical axis labelled $p$ indicates the Planck length) and growing to be equal to the Hubble radius  at time $t_R$ indicates the physical wavelength of the comoving scale which had Planck length at time $t_i$.  In the drawing we assume that the String Gas phase is quasi-static. In this case the physical length of a fixed comoving mode is constant in time.}
\end{center}
\end{figure*}
%\end{widetext}

For our model, we demand that scales which are currently probed in CMB experiments remain outside of the Hubble radius at all early times. Otherwise, the spectrum of cosmological fluctuations and gravitational waves produced in the early phase will be disrupted. This sets an upper bound on the duration of inflation. On the other hand, in order for inflation to solve the flatness problem, the period of inflation has to be sufficiently long. Below we show that these two conditions are inconsistent for simple slow-roll inflation, but that they can become consistent in the case of power law or warm inflation. 

To be specific, we will assume (as in \cite{Kamali1}) an inflationary phase during which the equation of state parameter $w$ is constant. This is obtained if
\be \label{kin}
{\dot{\varphi}}^2 \, = \, \beta V(\varphi) \, ,
\ee
where $\beta < 1$ is a constant. In this case the energy density in $\varphi$ scales as
\be
\rho(t) \, \sim \, a(t)^{-3\beta / (1 + \beta/2)} \, .
\ee
Such a dynamics can be obtained in cold inflation with an exponential potential \footnote{The potential needs to be modified at low values of the potential energy to allow for a {\it graceful exit} from inflation.} or in the context of warm inflation.

If we take $\Omega_K < 10^{-2}$ (where $\Omega_K$ is the fractional contribution of spatial curvature to the critical energy density) at the present time and we assume that the radiation phase of Standard Cosmology starts right after inflation, then the constraint on $\Omega_K$ at the end of inflation is
\be \label{OmegaBound}
\Omega_K(t_R) \, < \, 10^{-2} \frac{T_0T_{eq}}{T_R^2} \, ,
\ee
where $T_0$ is the present CMB temperature and $T_{eq}$ is the temperature at the time of equal matter and radiation. In order for inflation to solve the flatness problem, the relative decrease in the density of curvature during the inflationary phase must be greater than the increase which takes place after inflation. Thus, in order to be consistent with (\ref{OmegaBound}), we require
\be \label{cond1}
\bigl( \frac{a_i}{a_R} \bigr)^{2 - 3\beta / (1 + \beta /2)} \, < \, 10^{-2} \frac{T_0T_{eq}}{T_R^2} \, ,
\ee
where $a_i$ and $a_R$ denote the values of the scale factor at the beginning and the end of inflation, respectively.

On the other hand, we require that the comoving scale corresponding to the current Hubble radius remain larger than the Hubble length at the beginning of inflation, i.e.
\be \label{cond2a}
H(t_i)^{-1} \, < \, H_0^{-1} \frac{T_0}{T_R} \frac{a_i}{a_R} \, ,
\ee
where $H_0$ is the current Hubble radius, and $a_i$ and $a_R$ denote the scale factor at times $t_i$ and $t_R$, respectively. Making use of the Friedmann equation we have
\ba \label{help1}
H(t_i)^2 \, &=& \, \frac{1}{3} m_{pl}^{-2} (1 + \beta/2) V_i \\
H_0^2 \, &=& \, \frac{1}{3} m_{pl}^{-2} g^{*} T_0^3 T_{eq} \nonumber
\ea
where $V_i$ is the value of the scalar field potential energy at the beginning of inflation, and $g^{*}$ is the number of spin degrees of freedom in the radiation bath. Note that the factor $(1 + \beta/2)$ takes into account the kinetic energy which is not negligible unless $\beta \ll 1$. Assuming instantaneous onset of the radiation phase after inflation we can express $T_R$ in (\ref{cond2a}) in terms of the value of the potential energy $V_R$ at the end of inflation:
\be \label{help2}
(1 + \beta/2) V_R \, = \, g^{*}(T_R) T_R^4 \, .
\ee
Making use of (\ref{help1}) and (\ref{help2}), (\ref{cond2a}) becomes
\be \label{cond2b}
\frac{V_R}{V_i} \, < \, \frac{T_R^2}{T_0 T_{eq}} \bigl( \frac{a_i}{a_R} \bigr)^2 \, ,
\ee
if we neglect entropy production after inflation (in which case the $g^{*}$ factors are the same). Note that the factors of $1/3$ and $(1 + \beta/2)$ have cancelled out. In order to compare with the condition (\ref{cond1}) it is useful to rewrite (\ref{cond2b}) as
\be \label{cond2c}
\bigl( \frac{a_i}{a_R} \bigr)^2 \, > \, \frac{T_0T_{eq}}{T_R^2} \frac{V_R}{V_i} \, .
\ee

In the case of standard slow-roll inflation $\beta$ can be set to zero and the ratio of the potential energies at the beginning and end of inflation is approximately 1. Hence, the conditions (\ref{cond1}) and (\ref{cond2c}) are clearly in conflict. Thus, in this case inflation and String Gas Cosmology cannot be combined such that inflation solves the flatness problem while the cosmological fluctuations on large scales are given by String Gas Cosmology.

In order to determine whether the conditions (\ref{cond1}) and (\ref{cond2b}) can be consistent in a more general context, we must find an approximate expression for the ratio of the potential energies at the beginning and end of inflation. Following what was done in the Appendix of \cite{Kamali1}, we start with the evolution equation of $\varphi$ during inflation
\be \label{EoM}
\Gamma {\dot{\varphi}} \, = \, - V^{\prime} \, ,
\ee
where $\Gamma = 3H$ in the case of cold inflation, and $\Gamma > 3H$ is the friction coefficient in the case of warm inflation \footnote{Note that we are neglecting the acceleration term in the scalar field equation of motion. This approximation ceases to be justified when $\beta$ approaches the value 1.}. To be specific, we set
\be \label{Fdef}
\Gamma \, \equiv \, {\cal F} H \, ,
\ee
and we take ${\cal{F}}$ to be a constant. Multiplying both sides of (\ref{EoM}) by ${\dot{\varphi}}$, making use of (\ref{kin}) and integrating both sides yields
\be
\int \Gamma \beta dt \, = \, - {\rm{ln}} \frac{V_i}{V_R} \, .
\ee
Making use of (\ref{Fdef}) we obtain
\be \label{Ratio}
\frac{V_i}{V_R} \, = \, \bigl( \frac{a_R}{a_i} \bigr)^{{\cal F} \beta} \, .
\ee
In this case (\ref{cond2c}) becomes
\be \label{cond2d}
\bigl( \frac{a_i}{a_R} \bigr)^{2 - {\cal F}\beta}  \, > \, \frac{T_0T_{eq}}{T_R^2} \, ,
\ee
and conditions (\ref{cond1}) and (\ref{cond2d}) can be consistent provided that 
\be \label{Result}
{\cal F} \beta \, > \frac{3\beta}{1 + \beta/2} \, .
\ee
This is easy to achieve in the case of warm inflation when ${\cal{F}}$ is significantly larger than 3. The condition, however, can also be satisfied in cold inflation provided that $\beta$ is not too small.

Recall that the result (\ref{Ratio}) was obtained neglecting the ${\ddot{\varphi}}$ term in the scalar field equation of motion. For cold inflation (${\cal{F}} = 3$) with $\beta < 1$ 
this is a bad approximation. As shown in the Appendix of \cite{Kamali1}, the second derivative term can be explicitly taken into account and one obtains
\be \label{Ratio2}
\frac{V_i}{V_R} \, = \, \bigl( \frac{a_R}{a_i} \bigr)^{ 3 \beta / \sqrt{1 + \beta/2}} \, .
\ee
In this case, (\ref{cond2c}) becomes
\be \label{cond2e}
\bigl( \frac{a_i}{a_R} \bigr)^{2 - 3\beta / \sqrt{1 + \beta/2}}  \, > \, \frac{T_0T_{eq}}{T_R^2} \, ,
\ee
In this case conditions (\ref{cond1}) and (\ref{cond2e}) can be consistent. In fact, the larger the value of $\beta$, the easier it becomes to simultaneously satisfy both criteria.

\section{Consistency with the TCC}

We now show that the above scenario for combining String Gas Cosmology and inflation can be made consistent with the TCC. The TCC \cite{BeVa} states that no scales which were trans-Planckian at the beginning of inflation ever exited the Hubble horizon. The mathematical formulation of this condition is
\be \label{TCC}
\frac{a_R}{a_i} l_{pl} \, < \, H^{-1}(t_R) \, .
\ee
Taking the square of both sides, using the Friedmann equation to substitute for $H(t_R)$ in terms of $T_R$, and rearranging, this condition becomes
\be \label{TCC2}
\bigl( \frac{a_i}{a_R} \bigr)^2 \, > \, \frac{1}{3} g^{*}(T_R) \bigl( \frac{T_R}{m_{pl}} \bigr)^4 \, ,
\ee
which states that the period of inflation cannot last too long. 

However, in order for inflation to solve the flatness problem, there is a lower bound (\ref{cond1}) on the duration of inflation. This lower bound is consistent with the TCC if
\be
\frac{1}{3} g^{*}(T_R) \bigl( \frac{T_R}{m_{pl}} \bigr)^4 \, < \, \bigl( \frac{a_i}{a_R} \bigr)^2 \,
< \, \bigl( 10^{-2} \frac{T_0 T_{eq}}{T_R^2} \bigr)^{1 / (1 - 3 \beta / (2 + \beta))} \, ,
\ee
which can be consistent only if the following inequality
\be
\bigl[  \frac{1}{3} g^{*}(T_R) \bigr]^{1 - {\tilde{\beta}}} \bigl( \frac{T_R}{m_{pl}} \bigr)^{4(1 - {\tilde{\beta}})}
\, < \, 10^{-2} \frac{T_0T_{eq}}{T_R^2} \, .
\ee
is satisfied, where 
\be
{\tilde{\beta}} \, \equiv \, \frac{3\beta}{2 + \beta} \, .
\ee 
This can be written as the following upper bound on the temperature after the end of inflation (which is the energy scale at the end of inflation)
\ba
\frac{T_R}{m_{pl}} \, &<& \, \bigl[  3 g^{*}(T_R)^{-1} \bigr]^{(1 - {\tilde{\beta}}) / (6 - 4{\tilde{\beta}})} \\
& & \times 10^{-2 / (6 - 4{\tilde{\beta}})} \bigl( \frac{T_0T_{eq}}{m_{pl}^2} \bigr)^{1 / (6 - 4{\tilde{\beta}})} \, . 
\nonumber
\ea
This upper bound depends on the value of $\beta$. 

Taking the example $\beta = 1/2$ we obtain 
\be \label{Tvalue1}
\frac{T_R}{m_{pl}} \, < \, {\cal{O}}(1) 10^{-16} \, ,
\ee
and in the limit $\beta \rightarrow 0$
\be \label{Tvalue2}
\frac{T_R}{m_{pl}} \, < \, {\cal{O}}(1) 10^{-11} \, ,
\ee
which indicates that a very low energy scale at the end of inflation is required (significantly smaller than the bound obtained from the TCC alone for canonical inflation). 

Given the low temperature at the end of inflation we might worry whether our scenario is indeed able to solve the flatness problem. If the onset of inflation takes place at a temperature much lower than the Hagedorn temperature, then fine tuning of the initial curvature would be required in order to prevent the recollapse of the universe before inflation can set in. Thus, we must determine the energy scale when inflation begins. A lower bound on this energy scale can be obtained by making use of the lower bound (\ref{cond1}) on the change of the scale factor $a_i / a_R$ during the inflationary period, coupled to the equation (\ref{Ratio}) which relates the change in the scale factor to the change in the potential energy. To be specific, let us use the values $\beta = 1/2$, ${\cal{F}} = 3$ and the value of $T_R$ given by (\ref{Tvalue1}). In this case we find for the energy scale $\eta$ at the beginning of inflation
\be \label{scalebound}
\eta \, > \, 10^{-6} m_{pl} \, .
\ee

In order for string thermal fluctuations in the Hagedorn phase to have the right amplitude to explain the currently observed structures, the energy scale in the Hagedorn phase (which is the string scale) must be roughly $10^{-2} m_{pl}$ \cite{NBV, BNPV2}. In order for our proposed merger of string gas cosmology and power law inflation to work, we need the energy scale at the beginning of inflation to be lower than the energy scale in the string gas phase. As explained in the previous paragraph, we would like the energy scale at the onset of inflation to be only slightly lower than this scale.  As we see from (\ref{scalebound}), it is easy to satisfy this condition for the value of $\beta$ which we have chosen. In the limit $\beta \rightarrow 0$ it becomes more difficult to realize our scenario since the lower bound on the value of $\eta$ decreases and approaches $T_R$, and at the same time the upper bound on $\eta$ obtained from the TCC also decreases towards the value $T_R$.

\section{Cosmological Fluctuations}

In this section we discuss the spectrum of cosmological fluctuations which emerges from our model. We work in the context of the usual theory of cosmological perturbations \cite{MFB, RHBfluctsrev} and consider the perturbed metric in longitudinal gauge
\be
ds^2 \, = \, a^2(\eta) [(1 + 2\Phi) d\eta^2 - (1 - 2\Phi) d{\bf{x}}^2] \, ,
\ee
where $\Phi(x, t)$ is the relativistic generalization of the Newtonian gravitational potential, $\eta$ is conformal time and ${\bf{x}}$ are the comoving spatial coordinates (here we neglect the spatial curvature of the background). When during the inflationary phase the matter content is dominated by the scalar field $\varphi$, the matter field can be expanded in terms of its background value $\varphi_0(\eta)$ and the fluctuations $\delta \varphi({\bf{x}}, \eta)$. The canonical variable for the cosmological fluctuations is \cite{MukhSas}
\be
v \, = \, a \bigl( \delta \varphi + \frac{\varphi_0^{\prime}}{\cal{H}} \bigr) \Phi \, ,
\ee
where ${\cal{H}}$ is the Hubble expansion rate in comoving time, and a prime denotes a derivative with respect to comoving time. We need to compute the power spectrum of ${\cal{R}}$, the curvature fluctuation in comoving coordinates (coordinates in which $\delta \varphi = 0$). This is given by
\be
{\cal{R}} \, = \, z^{-1} v \, ,
\ee
where 
\be
z \, = \, a \frac{\varphi_0^{\prime}}{\cal{H}} \, .
\ee
The power spectrum of cosmological perturbations is defined via
\be
{\cal{P}}_{\cal{R}}(k) \, = \, k^3 |{\cal{R}}(k)|^2 \, ,
\ee
where ${\cal{R}}(k)$ is the Fourier mode of ${\cal{R}}$. The spectral slope $n_s$ is defined by
\be
{\cal{P}}_{\cal{R}}(k) \, \sim \, k^{n_s - 1} \, ,
\ee
and scale-invariance corresponds to $n_s = 1$.

The equation of motion for the Fourier mode $v_k$ of $v$ is
\be
v_k^{\prime \prime} + \bigl( k^2 - \frac{z^{\prime \prime}}{z} \bigr) v_k \, = \, 0 \, .
\ee
Since in general (except in the case of a radiation-dominated universe) $z^{\prime \prime}/z$ is proportional to ${\cal{H}}^2$, the equation implies that $v$ oscillates on sub-Hubble scales while it is squeezed on super-Hubble scales, with the dominant mode scaling as
\be
v_k \, \sim \, z \, 
\ee
(in an expanding background). This implies, in turn, that ${\cal{R}}$ is constant on super-Hubble scales (except during a phase transition of the background). 

As discussed in \cite{NBV, BNPV2}, thermal fluctuations of a gas of closed strings in the Hagedorn phase generates a scale-invariant spectrum of cosmological fluctuations with amplitude
\be \label{LSampl}
{\cal{P}}_{\cal{R}}(k) \, = \, B \bigl( \frac{\eta_s}{m_{pl}} \bigr)^4 \, \equiv \, {\cal{P}}_{SG}(k) \, ,
\ee
where $\eta_s$ is the string energy scale and $B$ is a constant of the order one. For a string scale similar to the scale of Grand Unification (a scale suggested in the original textbook \cite{GSW} on superstring theory), the amplitude of the resulting spectrum agrees with observations.

In our scenario the fluctuations on large scales are those set up in the Hagedorn phase. Since ${\cal{R}}$ is conserved on super-Hubble scales during the inflationary phase, the spectrum of ${\cal{R}}$ is unchanged on these scales, and given by (\ref{LSampl}).

There is a cutoff scale $k_1$ which corresponds to the mode whose wavelength equals the Hubble radius at the beginning of inflation and is given by
\be
k_1 \, = \, \frac{1}{\sqrt{3}} \frac{V_i^{1/2}}{m_{pl}} \bigl( \frac{V_R}{V_i} \bigr)^{1 / ({\cal{F}} \beta)}
(g^{*})^{1/4} \frac{T_0}{V_R^{1/4}} (1 + \beta/2)^{1/2} \, .
\ee
Modes with $k > k_1$ were sub-Hubble for a time interval $t_i < t < t_H(k)$, where $t_H(k)$ is the time when the scale exits the Hubble radius again during the inflationary phase. Since on sub-Hubble scales it is the variable $v$ whose amplitude is conserved, the amplitude of the cosmological perturbation ${\cal{R}}$ decreases as $z^{-1}$ during this time interval. Since we are working with a constant equation of state during the inflationary phase, we have $z \sim a$, and hence
\be \label{SGspectrum}
{\cal{P}}_{\cal{R}}(k) \, = \, \bigl( \frac{a(t_i)}{a(t_H(k))} \bigr)^2   {\cal{P}}_{SG}(k) \,\,\,\, k_2 > k > k_1 \, .
\ee
As shown in the Appendix, this is a very red spectrum with slope
\be \label{spectralindex}
n_s - 1 \, = \, - \frac{2 + \beta}{1 - \beta} \, .
\ee
As a consistency check, we note that in the limit $\beta \rightarrow 0$, the limit of almost exponential inflation, the spectral slope is $1 - n_s  \rightarrow -2$ which is expected since the spectrum during almost exponential inflation changes by $\delta n = - 2$.

In the case of inflation there is a second source of fluctuations, namely those produced in the inflationary phase. In the case of warm inflation it is the thermal fluctuations generated during the inflationary phase. There is a transition momentum scale $k_2$ above which the thermal fluctuations exceed those produced in the original Hagedorn phase. The spectrum of thermal fluctuations in warm inflation is given by \cite{warmflucts}
\be
{\cal{P}}_{\cal{R}}(k)^{\rm{warm}}  \, = \, \frac{\sqrt{3}}{4 \pi^{3/2}} \frac{H^3 T}{{\dot{\varphi}}^2} \bigl(\frac{Q}{Q_3} \bigr)^9 Q^{1/2} \, ,
\ee 
where ${\cal{F}} = 3Q$, and $Q_3 = 7.3$. The resulting spectral tilt is
\be
n_s - 1 \, = \, - \frac{1.5 \beta Q}{2 + \beta} \, ,
\ee
which corresponds to a slightly red spectrum. Note that in the limit $\beta \rightarrow 0$ we recover a scale-invariant spectrum. 

The value of $k_2$ is set by the condition
\be
{\cal{P}}_{\cal{R}}(k)^{\rm{warm}} \, = \,  \bigl( \frac{a(t_i)}{a(t_H(k))} \bigr)^2   {\cal{P}}_{SG}(k) \, ,
\ee
and a short calculation shows that it is given by the implicit equation
\be
\bigl( \frac{k_2}{k_1} \bigr)^2 \, = \, \frac{\eta_s^4}{V_i} \frac{V(t_H(k_2))^{1/4}}{m_{pl}} 
\ee
(up to factors of the order of 1), where $t_H(k)$ is the time when the mode $k$ exits the Hubble radius during the inflationary phase. As a consistency check, we note $k_2 > k_1$ for reasonable parameter values.

In the case of cold power law inflation, the quantum vacuum fluctuations at Hubble radius crossing have the spectrum
\be \label{inflspec}
{\cal{P}}_{\cal{R}}(k) \, \sim \, \bigl( \frac{H(t_H(k))}{m_{pl}} \bigr)^2 \, .
\ee
The tilt of this spectrum is a red one and is set by the value of $\beta$. For $\beta < 1$ the magnitude of the tilt is small. The transition scale $k_2$ is determined by equating (\ref{inflspec}) and (\ref{SGspectrum}), with the result
\be
k_2 \, \sim \, a(t_i) \frac{\eta_s^2}{m_{pl}} \, .
\ee

The shape of the resulting spectrum is sketched in Figure 2. The value of $k_2$ and the tilt for $k > k_2$ depend on whether we are considering cold power law or warm inflation.

%\begin{widetext}
\begin{figure*}[t]
\begin{center}
\includegraphics[scale=1]{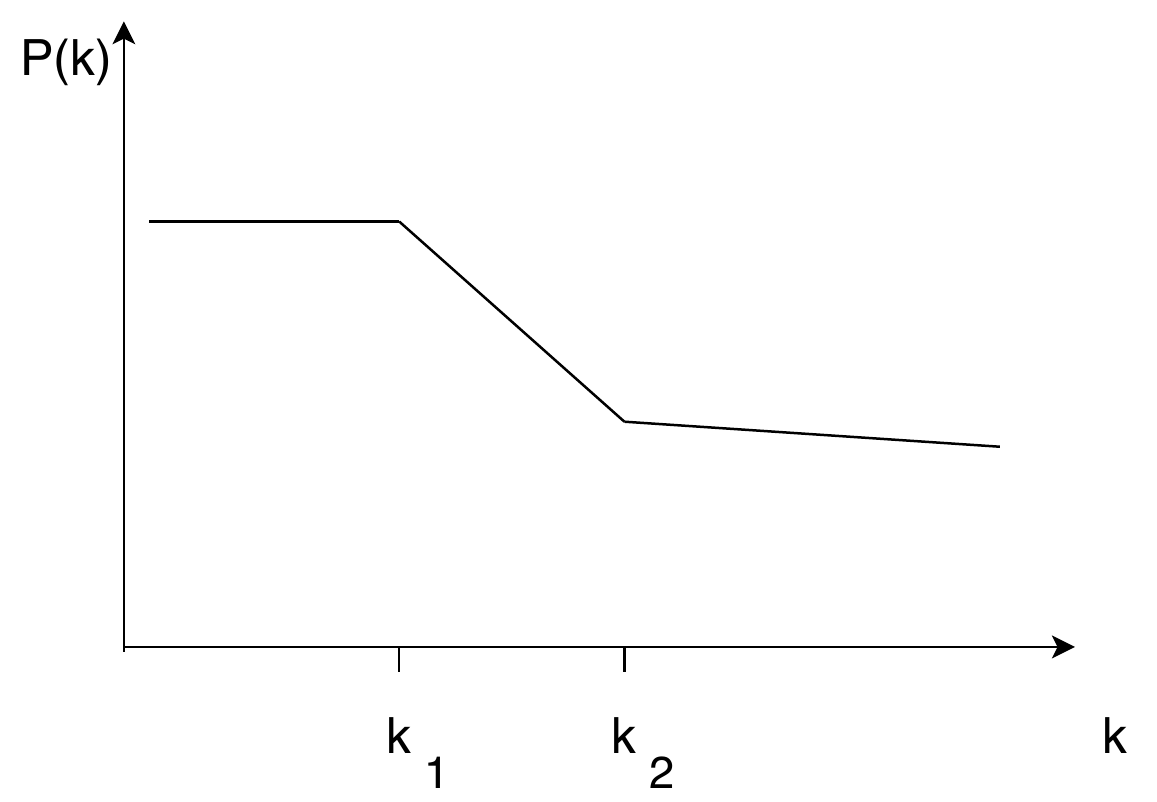}
\caption{Spectrum of cosmological perturbations in our model. The vertical axis is the dimensionless power spectrum, the horizontal axis is comoving wavenumber. On the largest scales $k < k_1$ the spectrum is the unperturbed string gas cosmology spectrum, on intermediate scales $k_1 < k < k_2$ the original string gas spectrum is damped during the time interval when the scale is inside the Hubble radius during the inflationary phase, but the fluctuations still dominate over those produced by warm inflation. Finally, for $k > k_2$ it is the fluctuations produced during inflation which dominate.}
\end{center}
\end{figure*}
%\end{widetext}

\section{Conclusions and Discussion} \label{conclusion}

We propose that the flatness problem of String Gas Cosmology can be solved by adding a post-Hagedorn phase of power law inflation. We have shown that there are parameter values for which the duration of inflation is sufficiently short such that fluctuation modes which are currently probed in CMB experiments were never inside the Hubble radius during inflation, while the observed spatial flatness can be obtained without tuning the initial spatial curvature during the early string phase. We have shown, in addition, that the model can be rendered consistent with the TCC. The phase of inflation which we are using can be either a phase of warm inflation or a phase of cold inflation with an exponential potential. Since the fluctuations generated during inflation are not those which are probed by current cosmological observations, the constraints on the inflationary model coming from demanding to obtain the right amplitude and slope of the spectrum of fluctuations disappear.

A prediction of our scenario is that the spectrum of cosmological fluctuations is not scale-invariant on small scales. On scales which were never inside the Hubble radius during inflation, the scale-invariant spectrum of cosmological perturbations produced in the string phase \cite{NBV} gets uniformly squeezed during inflation (i.e. squeezed by a factor which is independent of the wavenumber $k$). Note that here we are talking about what happens to the canonical variable $v$. On the other hand, modes which were sub-Hubble during inflation oscillate during inflation until they exit the Hubble radius. Hence, the squeezing factor decreases as $k$ increases, and the resulting spectrum obtains a tilt of $\Delta n_s \sim -2$ compared to the initial almost scale-invariant spectrum. The transition wave number $k_1$ can be tuned by adjusting the duration of inflation. On even smaller length scales ($k > k_2$), it is the fluctuations which are produced during the inflationary phase which dominate, and the spectral slope returns to an almost scale-invariant one (with a red tilt whose magnitude depends on the value of $\beta$).
\\

\section*{Acknowledgement}

\noindent The research at McGill is supported in part by funds from NSERC and from the Canada Research Chair program. The visit of VK to McGill has also been supported in part by the McGill Space Institute. RB acknowledges hospitality of the Institutes for Theoretical Physics and Particle Astrophysics of the ETH Zurich while this manuscript was being finalized for submission.

\section*{Appendix}

Here we derive the slope of the spectrum of fluctuations given in Eq. (\ref{spectralindex}). The starting point is Eq. (\ref{SGspectrum}). We need to determine the dependence on $k$ of $a(t_H(k))$. For a constant value of $\beta$, the equation of state parameter in the inflationary phase is
\be
w \, = \, - \frac{2 - \beta}{2 + \beta} \, ,
\ee
and the scale factor evolves in time as
\be
a(t) \, \sim \, t^{\alpha} \, ,
\ee
with
\be
\alpha \, = \, \frac{2 + \beta}{3 \beta} \, .
\ee
The Hubble radius crossing condition is
\be
k^{-1} a(t_H(k)) \, = \, t_H(k) \, ,
\ee
and hence
\ba
a(t_H(k))^{-2} \, &\sim& \, t_H(k)^{- 2 \alpha} \, \sim \, k^{ - 2 \alpha / (\alpha - 1)} \, \nonumber \\
&\sim& \, k^{-(2 + \beta) / (1 - \beta)} \, ,
\ea
which yields the result of Eq. (\ref{spectralindex}).

\end{document}